\documentclass{emulateapj}

\def\ms{m~s$^{-1}$}

\def\msun{$M_{\odot}$}
\def\mjup{$M_{\rm Jup}$}
\def\rjup{$R_{\rm Jup}$}
\def\rsun{$R_{\odot}$}

\def\feh{[M/H]}

\begin{document}

\title{A Smaller Radius for the Transiting Exoplanet WASP-10\lowercase{b}$^1$}    

\author{ John Asher Johnson\altaffilmark{2}, 
  Joshua N.~Winn\altaffilmark{2,3},
  Nicole E.~Cabrera\altaffilmark{4},
  Joshua A.~Carter\altaffilmark{3}
}

\email{johnjohn@ifa.hawaii.edu}

\altaffiltext{1}{Based on observations obtained with the University of
Hawaii 2.2~m telescope operated by the Institute for Astronomy.} 
\altaffiltext{2}{Institute for Astronomy, University
  of Hawaii, Honolulu, HI 96822; NSF Astronomy and Astrophysics
  Postdoctoral Fellow} 
\altaffiltext{3}{Department of Physics, and Kavli Institute for
  Astrophysics and Space Research, Massachusetts Institute  
  of Technology, Cambridge, MA 02139}
\altaffiltext{4}{Current Address: School of Physics, Georgia Institute of Technology,
  837 State Street, Atlanta, Georgia 30332-0430}

\begin{abstract}
  We present photometry of WASP-10 during a transit of its
  short--period Jovian planet. We employed the novel PSF--shaping
  capabilities of the OPTIC camera mounted on the UH 2.2\,m telescope to
  achieve a photometric precision of $4.7\times10^{-4}$ per 1.3~min
  sample. With this new light curve, in conjunction with stellar
  evolutionary models, we improve on existing measurements of the
  planetary, stellar and orbital parameters. We find a stellar radius
  $R_\star = 0.698 \pm 0.012$~\rsun\ and a planetary radius $R_P =
  1.080 \pm 0.020$~\rjup. The quoted errors do not include any
  possible systematic errors in the stellar evolutionary models. Our
  measurement improves the precision of the planet's radius by a
  factor of 4, and revises the previous estimate downward by 16\%
  (2.5$\sigma$, where $\sigma$ is the quadrature sum of the respective
  confidence limits). Our measured radius of WASP-10b is consistent
  with previously published theoretical radii for irradiated Jovian
  planets. 
\end{abstract}


\section{Introduction}

When an exoplanet transits its star it provides a valuable opportunity
to study the physical characteristics of planets outside our solar
system. Photometric survey teams are discovering a rapidly growing
number of short--period planets transiting relatively bright ($9
\lesssim V \lesssim 13$) Sun--like stars by monitoring the brightnesses
of hundreds of thousands of stars for weeks or months, and following
up on transit candidates with spectroscopy to rule out false
positives \citep{tres1,xo1,bakos07,wasp1,corot1}. With this windfall
of new planet discoveries comes a need 
for efficient and precise photometric follow--up. Measuring planet
properties precisely enough for meaningful tests of theoretical models
of planetary interiors requires transit light curves with good
photometric precision ($<$10$^{-3}$) and high cadence ($<$1~min). In
the past these requirements have 
often been met by combining photometry from many transit events
\citep{holman06, winn07c}, or observing from space using the
\emph{Hubble}\, and \emph{Spitzer}\, space telescopes \citep{brown01,
  gillon07b}. These approaches work but they are costly in time and/or
resources. 

Our approach is to capitalize on the strengths of a relatively new
type of detector: the orthogonal transfer array \citep[OTA;
][]{tonry97}. OTA detectors are charge-coupled devices (CCDs)
employing a novel four--phase gate structure to allow accumulated
charge to be shifted in two dimensions during an exposure. Usually,
this capability is used to compensate for image motion, providing
on--chip tip--tilt correction and a narrower point spread function
(PSF). However, as demonstrated by \citet{howell03}, OTAs can also be
utilized to deliberately broaden PSFs in a manner that is well suited
for high--precision photometry. The method of ``PSF shaping'' uses a
raster scan to distribute the accumulated signal from each star over
many pixels within a well-defined region of the detector, thereby
collecting more light per exposure and increasing the observing duty
cycle compared to traditional CCDs. This OTA observing mode provides
benefits similar to defocusing, but without the downside of a
spatially and chromatically varying PSF \citep{howell03, tonry05}.

In this Letter, we present new OTA-based photometry for a transiting
planetary system recently announced by the Super Wide-Angle Search for
Planets (SuperWASP) consortium. The planet, WASP-10b, is a 3.3~\mjup\
planet in an eccentric, 3.09~day orbit around a $V=12.7$ K--type dwarf
\citep[][hereafter C08]{wasp10}. C08 measured a radius for WASP-10b
that is much larger than predicted by models of planetary
interiors. In \S~\ref{data} we present our observations. We describe
our data reduction procedures in \S~\ref{reduction} and our light
curve modeling procedure and error analysis in \S~\ref{model}. We
present our results in \S~\ref{results} and conclude in
\S~\ref{discussion} with a brief discussion of our findings.

\section{Observations}
\label{data}

\begin{figure*}[!ht]
\epsscale{0.8}
\plotone{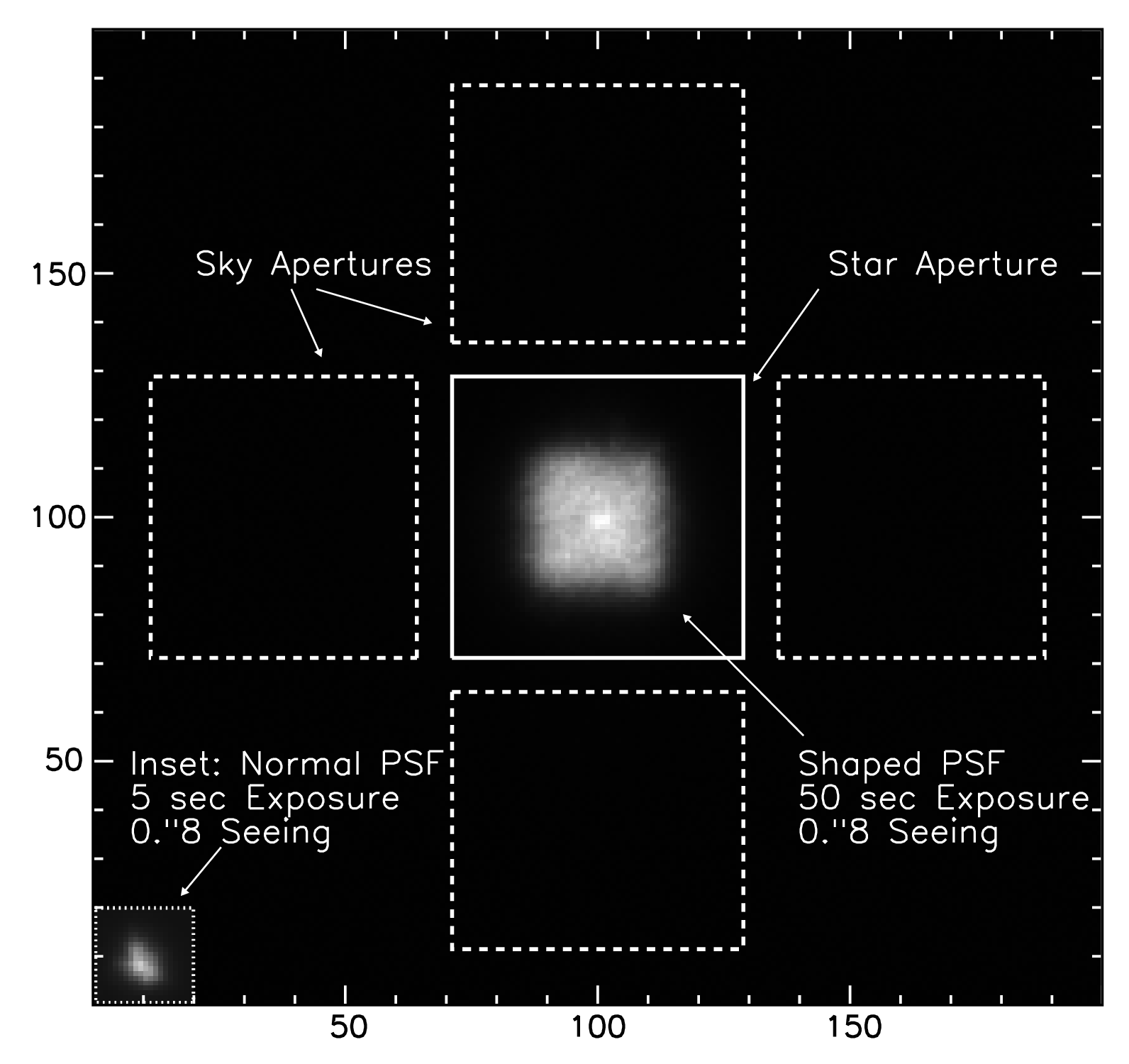}
\caption{Example of an OPTIC PSF selected from the WASP-10
  field. Throughout the exposure, the accumulated signal 
  from the incident stellar PSF is shifted into neighboring pixels 
  to produce a PSF confined to an approximately box--shaped
  region. The inset in the lower left region shows a separate, 5--second
  ``snap-shot'' exposure of WASP-10 with no OT guiding. The solid line
  shows the star aperture used to calculate the instrumental
  magnitude of the star. The background level is estimated from the
  mean level measured from the four flanking boxes  
  (dashed lines). \label{fig:psf}}   
\end{figure*}

We observed WASP-10 on UT August 16, spanning a transit time predicted
by the ephemeris of C08. We used the Orthogonal Parallel Transfer
Imaging Camera (OPTIC) on the University of Hawaii 2.2~m telescope
atop Mauna Kea, Hawaii. OPTIC consists of two $2048 \times 4096$ pixel
Lincoln Lab CCID28 OTA detectors with 15~$\mu$m pixels, corresponding
to 0\farcs135 pixel$^{-1}$ and a $9\farcm2\times9\farcm2$ field of
view \citep{howell03}. Each detector is read out through two
amplifiers, dividing the frame into four quadrants. The portions of
each quadrant closest to the amplifiers can be read out without
disturbing the rest of the detector. Small subregions near the
amplifiers can be read out rapidly (at a rate of up to 100 Hz) to
monitor the position of one to four guide stars. The centroids of the
guide stars are analyzed and can be used to shift charge in the
science regions to compensate for image motion due to seeing and
telescope guiding errors. In our application, we used a single OTA
guide star as a reference point to define the origin of the
shaped--PSF raster scan. We maximized our field of view by using the
remaining three guide regions to image the target field.

We observed WASP-10 continuously for 4.35~hr spanning the predicted
mid-transit time. Conditions were photometric. The seeing ranged from
0\farcs5 to 1\farcs2. We observed through a Sloan $z'$ filter, the
reddest broad--band filter available, to minimize the effects of
differential atmospheric extinction on the relative photometry, and to
minimize the effects of limb darkening on the transit light curve. We
shifted the charge during each integration to form a square PSF with a
side length of 20 pixels (2\farcs7). During each exposure, the
on--chip raster scan was traced out 11 times to uniformly fill in the
shaped--PSF region. An example of a WASP-10 PSF is shown in
Figure~\ref{fig:psf}. The integration time was 50~s, the readout time
was 25~s, and the reset and setup operations took an additional 4~s.
The resulting cadence was 79 seconds, corresponding to a 63\% duty
cycle.

\section{Data Reduction}
\label{reduction}

To calibrate the science images, we first subtracted the bias level
corresponding to each of the four amplifiers. Visual examination of a
median bias frame revealed that the bias offset is nearly constant
across the detector rows. We opted to estimate the bias level of the
individual science frames using the median of each 32--pixel--wide
over--scan row. For flat--field calibration, we normalized individual
dome flats by dividing each quadrant by the median value. We then
applied a median filter to the stacked set of normalized flats to form
a master flat for each night. Finally, we constructed a customized
flat for each science frame by convolving the unshifted master flat
with the OTA shift pattern (which is recorded in an auxiliary file at
the end of each exposure) \citep{howell03}.

We performed aperture photometry using custom procedures written in
IDL. The flux from each star was measured by summing the counts within
a square aperture centered on the PSF. Before the summation, we
subtracted the background level, taken to be the (outlier--rejected)
mean count level among four rectangular regions flanking the star
aperture. The star aperture and sky regions are illustrated in
Figure~\ref{fig:psf}.

\begin{figure*}[!ht]
\epsscale{1}
\plotone{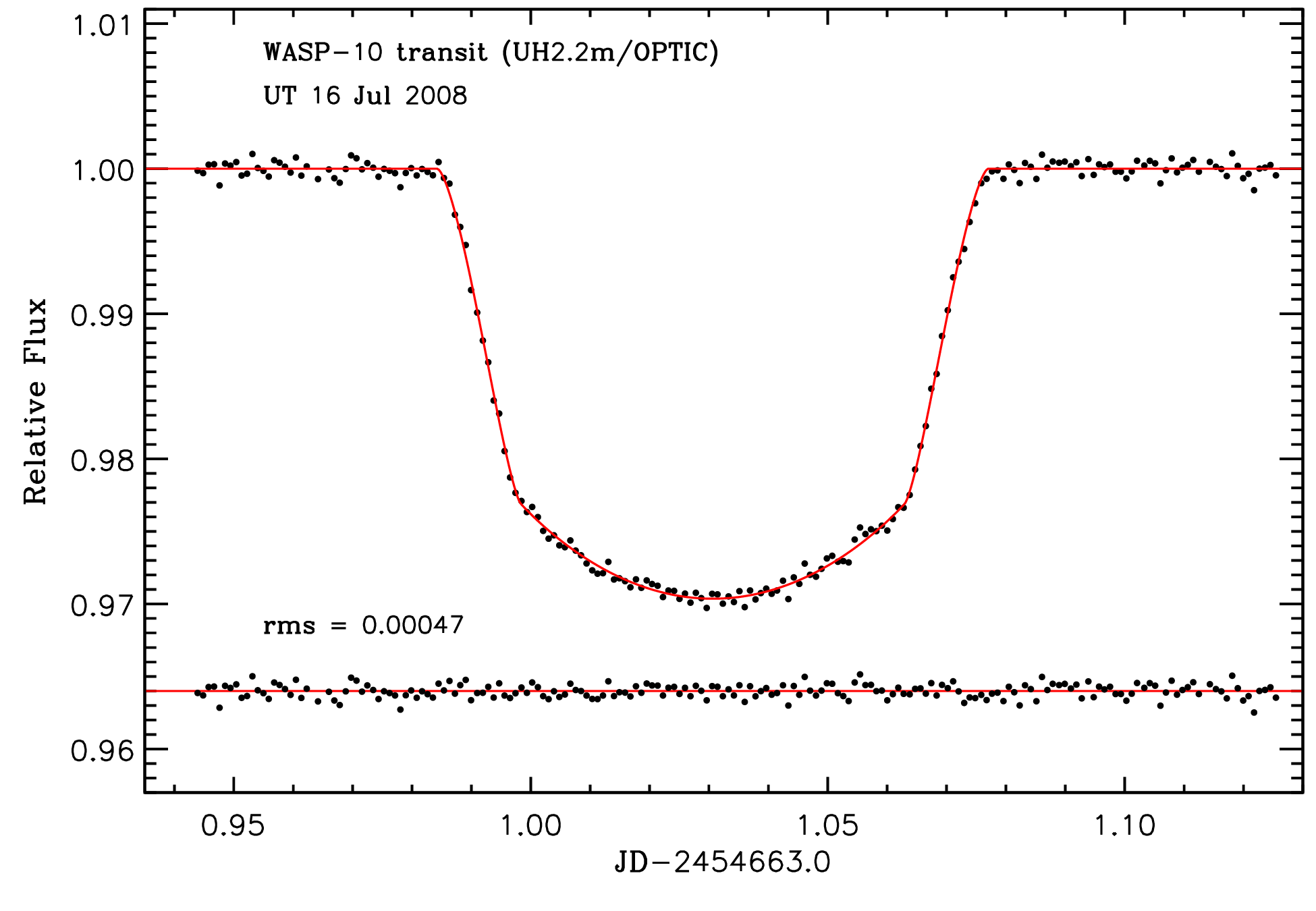}
\caption{Relative photometry of WASP-10. The solid red line
  shows the best--fitting model light curve and the residuals (O-C) are
  plotted belo.  \label{fig:lc1}}  
\end{figure*}

Ideally, the PSF shaping procedure confines the accumulated charge for
each star within a well--defined region of the detector that is common
to all images. However, the telescope focus gradually drifted,
resulting in variations in the full width at half maximum (FWHM) of
the shaped-PSFs ranging from the nominal 20 pixels up to 24 pixels
before we noticed this problem and refocused the telescope. We also
noticed a faint halo surrounding the main PSF, extending to several
pixels beyond the nominal PSF width. To ensure that our digital
aperture captures all of the incident flux, we used a very wide
aperture width of 50 pixels, chosen because it gave the smallest
scatter in relative flux outside of the transit. The sky regions
extend from 30 to 70 pixels from the center of the PSF.

To correct for variations in sky transparency, we divided the the flux
of WASP-10 by the sum of the fluxes of 5 comparison stars with
apparent magnitudes comparable to WASP-10.  We used the
root--mean--square (rms) scatter of the out--of--transit (OOT) points
as an estimate for the individual measurement uncertainties.  Our
relative photometric measurements of WASP-10 are shown in
Figure~\ref{fig:lc1}, and listed in Table~\ref{tab:WASP-10} along with
with the heliocentric Julian dates (HJD) of observation.  

\section{Modeling}
\label{model}

We determined the stellar, planetary and orbital characteristics by
(1) fitting the light curve with a model consisting of a star and
planet in Keplerian orbit around their center of mass; (2) combining
the results with stellar-evolutionary models to break the modeling
degeneracies inherent in the light-curve model.

\subsection{Photometric Model}

Our light curve fitting procedure is similar to that of
\citet{holman06} and \citet{winn07b}, and uses the analytic formulas
of \citet{mandelagol} to compute the integrated flux from the
uneclipsed stellar surface as a function of the relative positions of
the star and planet. The parameters were the stellar radius $R_*$ and
mass $M_*$, the planetary radius $R_P$ and mass $M_P$, the orbital
inclination $i$, period $P$, eccentricity $e$, argument of periastron
$\omega$, and mid--transit time $T_c$. We fixed $P$ at the value
$3.0927616$~d given by C08, which has negligible uncertainty for for
our purpose. Due to the well-known fitting degeneracies $M_p \propto
M_\star^{2/3}$ and $R_p \propto R_\star \propto M_\star^{1/3}$ (see,
e.g., Winn~2008) we
fixed $M_*$ at a fiducial value, thereby solving effectively for
$R_*/a$ and $R_P/a$ rather than $R_*$ and $R_P$. The planet mass $M_P$
is included for completeness but the data are highly insensitive to
this parameter. We measured $\omega$ and $e$ by fitting a Keplerian
model to the radial velocity (RV) measurements reported by C08 and fixed
the parameters in our subsequent analysis ($e=0.051$,
$\omega=153$~deg). Because the eccentricity is so close to zero, we
found that the actual values had very little effect on our
results. There were four other free parameters: two parameters
describing a linear function of airmass used to correct for
differential airmass extinction, and the two coefficients in the
quadratic limb darkening law.  We fitted the model to our photometric
observations using the goodness--of--fit statistic
\begin{equation}
\chi^2 = \sum_{i} \left(\frac{f_i({\rm obs}) - f_i({\rm calc})}{\sigma_i}\right)^2
\end{equation}
where $f_i({\rm obs})$ and $f_i({\rm calc})$ are the observed and computed fluxes,
respectively, and $\sigma_i$ is the uncertainty of each flux
measurement, set equal to 0.000475 for our measurements, for reasons
explained below.

To determine the best-fitting parameter values and their
uncertainties, we used a Markov Chain Monte Carlo algorithm with a
Gibbs sampler \citep[see e.g.][ and references therein]{winn08b}. We
created 10 chains of $10^6$ steps each, and combined them after
verifying that the chains were converged and well-mixed. The recorded
sequence of parameter combinations provides the posterior probability
distribution of each parameter. We record the 15.9, 50.0, and 84.1
percentile levels in the cumulative distribution. For each parameter
we quote the 50.0\% level (the median) as the ``best value,'' and use
the 84.1\% and 15.9\% levels to define the ``one sigma'' upper and
lower confidence limits.

After optimizing the model parameters, we calculated the residuals to
assess the photometric noise, and in particular the time-correlated
noise that may arise from instrumental, atmospheric and astrophysical
sources---the so--called ``red noise''---using the approach outlined
by \citet[][see also Gillon et al.~2006]{winn08b}. We averaged the
residuals into $M$ bins of $N$ points and measured the standard
deviation, $\sigma_N$, of the binned residuals. If the noise were
described by a Gaussian distribution then the rms of each grouping of
residuals should decrease as $N^{-1/2} [M/(M-1)]^{1/2}~\sigma_1$,
where $\sigma_1$ is the rms of the unbinned residuals. Correlated
noise would usually cause the observed scatter to be larger than this
expectation, by some factor $\beta$. We computed $\beta$ for binning
intervals ranging from 5 to 20 minutes, finding $\beta\approx 1$. For
this reason we set our measurement uncertainties equal to $\sigma_1 =
0.000475$ (0.5~mmag).

\subsection{Stellar Evolutionary Models}

To break the fitting degeneracies $M_p \propto M_\star^{2/3}$ and 
$R_p \propto R_\star \propto M_\star^{1/3}$, we required consistency
between the observed properties of the star, the stellar mean density
$\rho_\star$ that can be derived from the photometric parameter
$a/R_\star$ (Seager \& Mallen-Ornelas 2003, Sozzetti et al.~2007,
Torres et al. 2008), and
the Yonsei-Yale (Y$^2$) stellar evolution models by Yi et al.~(2001)
and Demarque et al.~(2004).  The inputs were $T_{\rm eff}=4675\pm
100$~K and [Fe/H]~$=0.03\pm 0.20$ from C08, and the stellar mean
density $\rho_\star = 3.099 \pm 0.088$~g~cm$^{-3}$ derived from the
light curve. We computed isochrones for the allowed range of
metallicities, and for stellar ages ranging from 0.1 to 14 Gyr. For
each stellar property (mass, radius, and age), we took a weighted
average of the points on each isochrone, in which the weights were
proportional to $\exp(-\chi^2_\star/2)$ with
\begin{equation}
\chi^2_\star =
\left[ \frac{\Delta{\rm [Fe/H]}}{\sigma_{{\rm [Fe/H]}}} \right]^2 +
\left[ \frac{\Delta T_{\rm eff}}{\sigma_{T_{\rm eff}}} \right]^2 +
\left[ \frac{\Delta \rho_\star}{\sigma_{\rho_\star}} \right]^2.
\end{equation}
Here, the $\Delta$ quantities denote the deviations between the
observed and calculated values at each point. The weights were further
multiplied by a factor taking into account the number density of stars
along each isochrone, assuming a Salpeter mass function. It is
important to keep in mind that this procedure assumes that any
systematic errors in the Y$^2$ isochrones are negligible.

\section{Results}
\label{results}

Table~\ref{tab:pars} lists the parameters for the WASP-10 planetary
system derived from the analysis of our new photometry. Of particular
interest are the the stellar and planetary radii, $R_* = 0.698 \pm
0.012$~\rsun\ and $R_P= 1.080 \pm 0.020$~\rjup, and the mid--transit
time, which is measured with a precision of 7~s. For completeness, we
also list the orbit parameters determined from our independent
analysis of the RV measurements of C08. Finally, we list several other
quantities derived from our light curves that can be used for
additional follow--up observations and modeling efforts.

Our measured planetary, stellar and orbital parameters differ
significantly from the values reported previously by C08. For example,
our measurement of the stellar radius is smaller by $1.9\sigma$, where
$\sigma$ is quadrature sum of the respective confidence limits, and
our measured orbital inclination is higher by $2.7\sigma$.  As for the
planetary radius, we refined the C08 value downward by $2.5\sigma$,
while improving the precision from $\approx0.08$~\rjup\ to
$0.02$~\rjup. The disagreement between the two results for the
planetary radius can be partly attributed to different stellar radii,
as well as different transit depths:
$(R_P/R_*)^2 = 0.02525^{+0.00024}_{-0.00028}$ from our analysis,
compared to $(R_P/R_*)^2 = 0.029 \pm 0.001$ from C08.

We do not know the reason for these discrepancies. C08 reported four
light curves: one based on the phased photometry from the 20~cm
SuperWASP survey telescope, and three higher precision light curves
observed with larger 1\,m class telescopes. The latter three data sets
were incomplete, covering ingress or egress but not both for a given
transit. In our experience, complete light curves such as ours are
much preferred, as they allow for more thorough checking of the noise
properties and any systematic effects. C08 did not report any tests
for red noise or other systematic effects, although such effects are
visually evident for the Tenagara light curve and may be present in
the other light curves as well. In addition, C08 did not report
whether they allowed the limb-darkening parameters to vary, or held
them fixed at values deemed appropriate; it has been shown by
Southworth~(2008) and others that holding them fixed can result in
underestimated parameter errors.

\section{Summary and Discussion}
\label{discussion}

We have presented high--precision photometry of the K--type star
WASP-10 during the transit of its close--in Jovian planet. Our
analysis has improved the precision with which the planetary, orbital and
stellar properties are known. From the photometry, we improved the
uncertainty in the orbit inclination angle by a factor of 2.7 and the
transit depth by a factor of 3.8, and we measured the transit mid-time
to 7 seconds. We also measured the stellar density ($\rho_*$)
from the transit light curve, which we used together with stellar
evolution models to measure the stellar mass to a precision of 4.6\%.  
We used the refined  stellar mass to improve the precision in the
planetary radius by a factor of 4. Notably, the 1.8\% uncertainty in
$R_P$ is dominated by the uncertainty in the stellar mass, rather than
the photometry. 

Our determination of the planetary radius represents a 16\% downward
revision of the value reported by C08. Those authors found that the
radius of WASP-10b was much larger than predicted by even a coreless
planet model. Following C08, we compared our revised radius of
WASP-10b to the theoretical predictions of \citet{fortney07}, who
provide tabulated radii for planets with a range of semimajor axes,
masses, ages and heavy-element core masses. For a hydrogen-helium
planet with properties appropriate to WASP-10, Fortney et al.\ predict
a radius of 1.13~\rjup\ for the case of a coreless planet with an age
of 4.5 Gyr. Adding 25, 50, and 100~$M_\oplus$ of heavy elements
(nominally in the form of a solid core) decreases the radius to 1.11,
1.09 and 1.06~\rjup, respectively.  The models are able to reproduce
our measured radius ($R_P = 1.08 \pm 0.02$~\rjup) by including a
heavy-element core with a mass ranging from 35~$M_\oplus$ to
95~$M_\oplus$.

We also compared our WASP-10b mass and radius measurements to the
theoretical models of \citet{bodenheimer03}, who predicted the radii
of gas giant planets as a function of mass, age and equilibrium
temperature. Assuming a planet temperature $T_{\rm eq} = T_{\rm
  eff}(R_*/a)^{1/2} = 1370$ (Table~2), \citet{bodenheimer03} predict
$R_P = 1.11$~\rjup\ for a 4.5 Gyr, solar-composition planet. The
predicted radius is essentially the same whether the composition is
taken to be purely solar or if a 40~$M_\oplus$ core of heavy elements
is included. It is also fairly insensitive to the value of $T_{\rm
  eq}$. Thus our measured radius is only 3\% ($1.5\sigma$) smaller
than predicted by \citet{bodenheimer03}.

We conclude that WASP-10b is not significantly ``inflated'' beyond the
expectations of standard models of gas giant planets. Instead, we find
that WASP-10b is among the densest exoplanets so far discovered. Its
mass and radius similar to another transiting planet, HD\,17156b (Winn
et al.\ 2008c). However, WASP-10 and HD\,17156 are very different host
stars. Whereas HD\,17156 is metal-rich ([Fe/H]$ = +0.24$) and has a
mass of 1.26~\msun, WASP-10 is a 0.75~\msun\ K dwarf with a much lower
atmospheric abundance ([M/H]$ = +0.03 \pm 0.2$; C08). WASP-10 stands
out as one of only nine known low-mass ($M_* < 0.8$~\msun) planet host
stars within 200~pc, and its planet is among the most massive so far
detected in this sample of low-mass
stars\footnote{http://exoplanets.org}.
 
We have found the OPTIC instrument to be capable of delivering
single-transit photometry that is precise enough for meaningful
comparisons of observed planetary properties and theoretical
models. Our WASP-10 light curve has a precision of $4.7\times10^{-4}$
per 1.3~min sample, or $6.8\times10^{-4}$ per 1~min bin, for a $V=12.7$ 
star (Figure~\ref{fig:lc1}). This is
comparable to the precision that has been reported for composite light
curves, based on observations of many transits, with traditional CCD
cameras. For example, \citet{johnson08b} presented a composite
$z$-band light curve for HAT-P-1 ($V=10.4$) based on observations of 8
transits, with a precision of $5.7\times10^{-4}$ in 1~min bins. OTAs
may also prove useful for the future detection and characterization of
transiting Neptune--mass and ``super--Earth'' planets.

\acknowledgements We gratefully acknowledge the assistance of the UH
2.2~m telescope staff, including Edwin Sousa, Jon Archambeau, Dan
Birchall, John Dvorak and Hubert Yamada. We are particularly grateful
to John Tonry for his clear and comprehensive instrument
documentation, and for making his previous OPTIC data available to us
in preparation for our observing run. JAJ is an NSF Astronomy and
Astrophysics Postdoctoral Fellow with support from the NSF grant
AST-0702821. NEC's research was supported by the University of Hawaii
Institute for Astronomy Research Experiences for Undergraduates (REU)
Program, which is funded by the National Science Foundation through
the grant AST-0757887. The authors wish to extend special thanks to
those of Hawaiian ancestry on whose sacred mountain of Mauna Kea we
are privileged to be guests. Without their generous hospitality, the
observations presented herein would not have been possible.


\begin{thebibliography}{22}
\expandafter\ifx\csname natexlab\endcsname\relax\def\natexlab#1{#1}\fi

\bibitem[{{Alonso} {et~al.}(2004){Alonso}, {Brown}, {Torres}, {Latham},
  {Sozzetti}, {Mandushev}, {Belmonte}, {Charbonneau}, {Deeg}, {Dunham}, {O'Donovan}, \& {Stefanik}}]{tres1}
{Alonso}, R., {Brown}, T.~M., {Torres}, G., {Latham}, D.~W., {Sozzetti}, A.,
  {Mandushev}, G., {Belmonte}, J.~A., {Charbonneau}, D., {Deeg}, H.~J.,
  {Dunham}, E.~W., {O'Donovan}, F.~T., \& {Stefanik}, R.~P. 2004, \apjl, 613,
  L153

\bibitem[{{Bakos} {et~al.}(2007){Bakos}, {Noyes}, {Kov{\'a}cs}, {Latham},
  {Sasselov}, {Torres}, {Fischer}, {Stefanik}, {Sato}, {Johnson}, {P{\'a}l},
  {Marcy}, {Butler}, {Esquerdo}, {Stanek}, {L{\'a}z{\'a}r}, {Papp}, {S{\'a}ri},
  \& {Sip{\H o}cz}}]{bakos07}
{Bakos}, G.~{\'A}., {Noyes}, R.~W., {Kov{\'a}cs}, G., {Latham}, D.~W.,
  {Sasselov}, D.~D., {Torres}, G., {Fischer}, D.~A., {Stefanik}, R.~P., {Sato},
  B., {Johnson}, J.~A., {P{\'a}l}, A., {Marcy}, G.~W., {Butler}, R.~P.,
  {Esquerdo}, G.~A., {Stanek}, K.~Z., {L{\'a}z{\'a}r}, J., {Papp}, I.,
  {S{\'a}ri}, P., \& {Sip{\H o}cz}, B. 2007, \apj, 656, 552

\bibitem[{{Barge} {et~al.}(2008){Barge}, {Baglin}, {Auvergne}, {Rauer},
  {L{\'e}ger}, {Schneider}, {Pont}, {Aigrain}, {Almenara}, {Alonso},
  {Barbieri}, {Bord{\'e}}, {Bouchy}, {Deeg}, {La Reza}, {Deleuil}, {Dvorak},
  {Erikson}, {Fridlund}, {Gillon}, {Gondoin}, {Guillot}, {Hatzes}, {Hebrard},
  {Jorda}, {Kabath}, {Lammer}, {Llebaria}, {Loeillet}, {Magain}, {Mazeh},
  {Moutou}, {Ollivier}, {P{\"a}tzold}, {Queloz}, {Rouan}, {Shporer}, \&
  {Wuchterl}}]{corot1}
{Barge}, P., {Baglin}, A., {Auvergne}, M., {Rauer}, H., {L{\'e}ger}, A.,
  {Schneider}, J., {Pont}, F., {Aigrain}, S., {Almenara}, J.-M., {Alonso}, R.,
  {Barbieri}, M., {Bord{\'e}}, P., {Bouchy}, F., {Deeg}, H.~J., {La Reza}, D.,
  {Deleuil}, M., {Dvorak}, R., {Erikson}, A., {Fridlund}, M., {Gillon}, M.,
  {Gondoin}, P., {Guillot}, T., {Hatzes}, A., {Hebrard}, G., {Jorda}, L.,
  {Kabath}, P., {Lammer}, H., {Llebaria}, A., {Loeillet}, B., {Magain}, P.,
  {Mazeh}, T., {Moutou}, C., {Ollivier}, M., {P{\"a}tzold}, M., {Queloz}, D.,
  {Rouan}, D., {Shporer}, A., \& {Wuchterl}, G. 2008, \aap, 482, L17

\bibitem[{{Bodenheimer} {et~al.}(2003){Bodenheimer}, {Laughlin}, \&
  {Lin}}]{bodenheimer03}
{Bodenheimer}, P., {Laughlin}, G., \& {Lin}, D.~N.~C. 2003, \apj, 592, 555

\bibitem[{{Brown} {et~al.}(2001){Brown}, {Charbonneau}, {Gilliland}, {Noyes},
  \& {Burrows}}]{brown01}
{Brown}, T.~M., {Charbonneau}, D., {Gilliland}, R.~L., {Noyes}, R.~W., \&
  {Burrows}, A. 2001, \apj, 552, 699

\bibitem[{{Cameron} {et~al.}(2007){Cameron}, {Bouchy}, {H{\'e}brard}, {Maxted},
  {Pollacco}, {Pont}, {Skillen}, {Smalley}, {Street}, {West}, {Wilson},
  {Aigrain}, {Christian}, {Clarkson}, {Enoch}, {Evans}, {Fitzsimmons},
  {Fleenor}, {Gillon}, {Haswell}, {Hebb}, {Hellier}, {Hodgkin}, {Horne},
  {Irwin}, {Kane}, {Keenan}, {Loeillet}, {Lister}, {Mayor}, {Moutou}, {Norton},
  {Osborne}, {Parley}, {Queloz}, {Ryans}, {Triaud}, {Udry}, \&
  {Wheatley}}]{wasp1}
{Cameron}, A.~C., {Bouchy}, F., {H{\'e}brard}, G., {Maxted}, P., {Pollacco},
  D., {Pont}, F., {Skillen}, I., {Smalley}, B., {Street}, R.~A., {West}, R.~G.,
  {Wilson}, D.~M., {Aigrain}, S., {Christian}, D.~J., {Clarkson}, W.~I.,
  {Enoch}, B., {Evans}, A., {Fitzsimmons}, A., {Fleenor}, M., {Gillon}, M.,
  {Haswell}, C.~A., {Hebb}, L., {Hellier}, C., {Hodgkin}, S.~T., {Horne}, K.,
  {Irwin}, J., {Kane}, S.~R., {Keenan}, F.~P., {Loeillet}, B., {Lister}, T.~A.,
  {Mayor}, M., {Moutou}, C., {Norton}, A.~J., {Osborne}, J., {Parley}, N.,
  {Queloz}, D., {Ryans}, R., {Triaud}, A.~H.~M.~J., {Udry}, S., \& {Wheatley},
  P.~J. 2007, \mnras, 375, 951

\bibitem[{{Christian} {et~al.}(2008){Christian}, {Gibson}, {Simpson}, {Street},
  {Skillen}, {Pollacco}, {Collier Cameron}, {Stempels}, {Haswell}, {Horne},
  {Joshi}, {Keenan}, {Anderson}, {Bentley}, {Bouchy}, {Clarkson}, {Enoch},
  {Hebb}, {H{\'e}brard}, {Hellier}, {Irwin}, {Kane}, {Lister}, {Loeillet},
  {Maxted}, {Mayor}, {McDonald}, {Moutou}, {Norton}, {Parley}, {Pont},
  {Queloz}, {Ryans}, {Smalley}, {Smith}, {Todd}, {Udry}, {West}, {Wheatley}, \&
  {Wilson}}]{wasp10}
{Christian}, D.~J., {Gibson}, N.~P., {Simpson}, E.~K., {Street}, R.~A.,
  {Skillen}, I., {Pollacco}, D., {Collier Cameron}, A., {Stempels}, H.~C.,
  {Haswell}, C.~A., {Horne}, K., {Joshi}, Y.~C., {Keenan}, F.~P., {Anderson},
  D.~R., {Bentley}, S., {Bouchy}, F., {Clarkson}, W.~I., {Enoch}, B., {Hebb},
  L., {H{\'e}brard}, G., {Hellier}, C., {Irwin}, J., {Kane}, S.~R., {Lister},
  T.~A., {Loeillet}, B., {Maxted}, P., {Mayor}, M., {McDonald}, I., {Moutou},
  C., {Norton}, A.~J., {Parley}, N., {Pont}, F., {Queloz}, D., {Ryans}, R.,
  {Smalley}, B., {Smith}, A.~M.~S., {Todd}, I., {Udry}, S., {West}, R.~G.,
  {Wheatley}, P.~J., \& {Wilson}, D.~M. 2008, ArXiv: 0806.1482, 806

\bibitem[{{Fischer} {et~al.}(2005){Fischer}, {Laughlin}, {Butler}, {Marcy},
  {Johnson}, {Henry}, {Valenti}, {Vogt}, {Ammons}, {Robinson}, {Spear},
  {Strader}, {Driscoll}, {Fuller}, {Johnson}, {Manrao}, {McCarthy},
  {Mu{\~n}oz}, {Tah}, {Wright}, {Ida}, {Sato}, {Toyota}, \&
  {Minniti}}]{fischer05a}
{Fischer}, D.~A., {Laughlin}, G., {Butler}, P., {Marcy}, G., {Johnson}, J.,
  {Henry}, G., {Valenti}, J., {Vogt}, S., {Ammons}, M., {Robinson}, S.,
  {Spear}, G., {Strader}, J., {Driscoll}, P., {Fuller}, A., {Johnson}, T.,
  {Manrao}, E., {McCarthy}, C., {Mu{\~n}oz}, M., {Tah}, K.~L., {Wright}, J.,
  {Ida}, S., {Sato}, B., {Toyota}, E., \& {Minniti}, D. 2005, \apj, 620, 481

\bibitem[{{Fortney} {et~al.}(2007){Fortney}, {Marley}, \& {Barnes}}]{fortney07}
{Fortney}, J.~J., {Marley}, M.~S., \& {Barnes}, J.~W. 2007, \apj, 659, 1661

\bibitem[{{Gillon} {et~al.}(2007){Gillon}, {Demory}, {Barman}, {Bonfils},
  {Mazeh}, {Pont}, {Udry}, {Mayor}, \& {Queloz}}]{gillon07b}
{Gillon}, M., {Demory}, B.-O., {Barman}, T., {Bonfils}, X., {Mazeh}, T.,
  {Pont}, F., {Udry}, S., {Mayor}, M., \& {Queloz}, D. 2007, \aap, 471, L51

\bibitem[{{Holman} {et~al.}(2006){Holman}, {Winn}, {Latham}, {O'Donovan},
  {Charbonneau}, {Bakos}, {Esquerdo}, {Hergenrother}, {Everett}, \&
  {P{\'a}l}}]{holman06}
{Holman}, M.~J., {Winn}, J.~N., {Latham}, D.~W., {O'Donovan}, F.~T.,
  {Charbonneau}, D., {Bakos}, G.~A., {Esquerdo}, G.~A., {Hergenrother}, C.,
  {Everett}, M.~E., \& {P{\'a}l}, A. 2006, \apj, 652, 1715

\bibitem[{{Howell} {et~al.}(2003){Howell}, {Everett}, {Tonry}, {Pickles}, \&
  {Dain}}]{howell03}
{Howell}, S.~B., {Everett}, M.~E., {Tonry}, J.~L., {Pickles}, A., \& {Dain}, C.
  2003, \pasp, 115, 1340

\bibitem[{{Johnson} {et~al.}(2008){Johnson}, {Winn}, {Narita}, {Enya},
  {Williams}, {Marcy}, {Sato}, {Ohta}, {Taruya}, {Suto}, {Turner}, {Bakos},
  {Butler}, {Vogt}, {Aoki}, {Tamura}, {Yamada}, {Yoshii}, \&
  {Hidas}}]{johnson08b}
{Johnson}, J.~A., {Winn}, J.~N., {Narita}, N., {Enya}, K., {Williams},
  P.~K.~G., {Marcy}, G.~W., {Sato}, B., {Ohta}, Y., {Taruya}, A., {Suto}, Y.,
  {Turner}, E.~L., {Bakos}, G., {Butler}, R.~P., {Vogt}, S.~S., {Aoki}, W.,
  {Tamura}, M., {Yamada}, T., {Yoshii}, Y., \& {Hidas}, M. 2008, ArXiv:
  0806.1734, 806

\bibitem[{{Mandel} \& {Agol}(2002)}]{mandelagol}
{Mandel}, K. \& {Agol}, E. 2002, \apjl, 580, L171

\bibitem[{{McCullough} {et~al.}(2006){McCullough}, {Stys}, {Valenti},
  {Johns-Krull}, {Janes}, {Heasley}, {Bye}, {Dodd}, {Fleming}, {Pinnick},
  {Bissinger}, {Gary}, {Howell}, \& {Vanmunster}}]{xo1}
{McCullough}, P.~R., {Stys}, J.~E., {Valenti}, J.~A., {Johns-Krull}, C.~M.,
  {Janes}, K.~A., {Heasley}, J.~N., {Bye}, B.~A., {Dodd}, C., {Fleming}, S.~W.,
  {Pinnick}, A., {Bissinger}, R., {Gary}, B.~L., {Howell}, P.~J., \&
  {Vanmunster}, T. 2006, \apj, 648, 1228

\bibitem[{{Tonry} {et~al.}(1997){Tonry}, {Burke}, \& {Schechter}}]{tonry97}
{Tonry}, J., {Burke}, B.~E., \& {Schechter}, P.~L. 1997, \pasp, 109, 1154

\bibitem[{{Tonry} {et~al.}(2005){Tonry}, {Howell}, {Everett}, {Rodney},
  {Willman}, \& {VanOutryve}}]{tonry05}
{Tonry}, J.~L., {Howell}, S.~B., {Everett}, M.~E., {Rodney}, S.~A., {Willman},
  M., \& {VanOutryve}, C. 2005, \pasp, 117, 281

\bibitem[{{Torres} {et~al.}(2008){Torres}, {Winn}, \& {Holman}}]{torres08}
{Torres}, G., {Winn}, J.~N., \& {Holman}, M.~J. 2008, \apj, 677, 1324

\bibitem[{{Winn} {et~al.}(2007{\natexlab{a}}){Winn}, {Holman}, {Bakos},
  {P{\'a}l}, {Johnson}, {Williams}, {Shporer}, {Mazeh}, {Fernandez}, {Latham},
  \& {Gillon}}]{winn07b}
{Winn}, J.~N., {Holman}, M.~J., {Bakos}, G.~{\'A}., {P{\'a}l}, A., {Johnson},
  J.~A., {Williams}, P.~K.~G., {Shporer}, A., {Mazeh}, T., {Fernandez}, J.,
  {Latham}, D.~W., \& {Gillon}, M. 2007{\natexlab{a}}, \aj, 134, 1707

\bibitem[{{Winn} {et~al.}(2007{\natexlab{b}}){Winn}, {Holman}, \&
  {Roussanova}}]{winn07c}
{Winn}, J.~N., {Holman}, M.~J., \& {Roussanova}, A. 2007{\natexlab{b}}, \apj,
  657, 1098

\bibitem[Winn(2008)]{2008arXiv0807.4929W} Winn, J.~N.\ 2008, arXiv:0807.4929 

\bibitem[{{Winn} {et~al.}(2008{\natexlab{b}}){Winn}, {Holman}, {Torres},
  {McCullough}, {Johns-Krull}, {Latham}, {Shporer}, {Mazeh}, {Garcia-Melendo},
  {Foote}, {Esquerdo}, \& {Everett}}]{winn08b}
{Winn}, J.~N., {Holman}, M.~J., {Torres}, G., {McCullough}, P., {Johns-Krull},
  C.~M., {Latham}, D.~W., {Shporer}, A., {Mazeh}, T., {Garcia-Melendo}, E.,
  {Foote}, C., {Esquerdo}, G., \& {Everett}, M. 2008{\natexlab{b}},
  ArXiv:0804.4475, 804

\end{thebibliography}

\clearpage
\begin{deluxetable}{ll}
\tablecaption{Relative Photometry for WASP-10\label{tab:WASP-10}}
\tablewidth{0pt}
\tablehead{
\colhead{Heliocentric Julian Date} &
\colhead{Relative Flux} 
}
\startdata
2454663.94419 & 0.99986 \\
2454663.94512 & 0.99970 \\
2454663.94604 & 1.00028 \\
2454663.94696 & 1.00031 \\
2454663.94788 & 0.99885 \\
... & ...
\enddata
\tablecomments{The full version of this table is available in the
  online edition, or by request to the authors.}
\end{deluxetable}

\begin{deluxetable}{lccc}
\tablecaption{System Parameters of WASP-10\label{tab:pars}}
\tablewidth{0pt}
\tablehead{
  \colhead{Parameter} & 
  \colhead{Value}     &
  \colhead{68.3\% Confidence Interval}     &
  \colhead{Comment}   \\
}
\startdata
\emph{Transit Parameters} & & \\
Mid-transit time, $T_c$~[HJD] & 2454664.030913 &$\pm 0.000082$ & A \\
Orbital Period, $P$~[days] & 3.0927616 & $\pm0.0000112$ & D \\
Planet-to-star radius ratio, $R_P/R_*$ & 0.15918 & $-0.00115, +0.00050$ & A \\
Planet-star area ratio, $(R_P/R_*)^2$ & 0.02525 & $-0.00028, +0.00024$ & A \\ 
Scaled semimajor axis, $a/R_*$  & 11.65 & $-0.13, +0.09$ & A \\
Orbit inclination, $i$~[deg] & 88.49 & $-0.17, +0.22$ & A \\
Transit impact parameter, $b$ & 0.299 & $-0.043, +0.029$ & A \\
Transit duration [hr] & 2.2271 & $-0.0068, +0.0078$ & A \\
Transit ingress or egress duration [hr] & 0.3306 & $-0.0075, +0.0098$ & A \\

 & & \\
\emph{Other Orbital Parameters} & & \\
Semimajor axis~[AU] & 0.03781 &$-0.00047$, $+0.00067$ & B \\
$e\cos{\omega}$ & -0.0453 & $\pm 0.02$ & C \\
$e\sin{\omega}$ & 0.0228 & $\pm 0.03$ & C \\
Velocity semiamplitude $K_*$~[\ms] &  533.1 &$ \pm 7.5$ & C \\
 & & \\
\emph{Stellar Parameters} & & \\
$M_*$~[$M_\odot$] & 0.75 & $-0.028,+0.040$ & B \\
$R_*$~[$R_\odot$] & 0.698 & $\pm 0.012$ & B \\
$\rho_*$~[$\rho_\odot$] & 3.099 & $\pm 0.088$ & A \\
$\log g_*$~[cgs]\tablenotemark{a} & 4.627 & $-0.0093$,$+0.0101$ & B \\
\feh & 0.03 & $\pm 0.2$ & D \\
$T_{\rm eff}$~[K] & 4675 & $\pm 100$ & D \\
 & & \\
\emph{Planetary Parameters} & & \\
$M_P$~[$M_{Jup}$] & 3.15  &   $-0.11$, $+0.13$  & B,C \\
$R_P$~[$R_{Jup}$] & 1.080 &   $\pm 0.020$  & B \\
Mean density, $\rho_P$~[$\rho_{Jup}$] & 3.11 & $\pm 0.20$ & B,C \\
$\log g_P$~[cgs] & 3.828 & $\pm 0.012$ & A \\
Equlibrium Temperature $T_{\rm eff}(R_*/a)^{1/2}$~[K] & 1370 & $\pm50$ & D 
\enddata

\tablecomments{Note.---(A) Determined from the parametric fit to our
  light curve. (B) Based on group A parameters supplemented by the
  Y$^2$ stellar evolutionary models. (C) Based on our analysis of the
  C08 RV measurements. (D) Reproduced from C08.}

\tablenotetext{a}{The $\log g_*$ in the table is the value implied by
  the $Y^2$ stellar evolution models, given the measured values of
  $T_{\rm eff}$, $[M/H]$, and $\rho_*$. It is in near agreement with
  the value $4.40\pm 0.20$ reported by C08, based on the widths of
  pressure-sensitive absorption lines in the stellar spectrum.}
\end{deluxetable}

\end{document}